\documentstyle[floats,twocolumn,prb,aps,epsfig]{revtex}
\begin{document}
\draft
\preprint{}
\title{Dynamically Induced Multi-Channel Kondo Effect}
\author{Tatsuya Fujii and Norio Kawakami} 
\address{Department of Applied Physics,
Osaka University, Suita, Osaka 565, Japan} 
\date{\today}
\maketitle
%%%%%%%%%%%%%%%                                          %%%%%%%%%%%%%%%%
%%%%%%%%%%%%%%%                ABSTRACT                  %%%%%%%%%%%%%%%%
%%%%%%%%%%%%%%%                                          %%%%%%%%%%%%%%%%
\begin{abstract}
We study how the multi-channel Kondo effect 
is dynamically induced to affect 
the photoemission and the inverse photoemission spectrum 
when an electron is emitted from (or added to)
the completely screened Kondo impurity with spin $S>1/2$. 
The spectrum thereby shows a power-law edge singularity
characteristic of the multi-channel Kondo model.
We discuss this anomalous behavior by using 
the exact solution of the multi-channel Kondo model and 
boundary conformal field  theory. 
The idea is further applied to the photoemission 
for quantum spin systems, in which the edge singularity is 
controlled by the dynamically induced 
overscreening effect with a {\it mobile} Kondo impurity.
\end{abstract}

%%%%%%%%%%%%%%%                                            %%%%%%%%%%%%%%%
%%%%%%%%%%%%%%%                                            %%%%%%%%%%%%%%%
%%%%%%%%%%%%%%%--------------------------------------------%%%%%%%%%%%%%%%
%%%%%%%%%%%%%%%        section 1 INTRODUCTION              %%%%%%%%%%%%%%%
%%%%%%%%%%%%%%%--------------------------------------------%%%%%%%%%%%%%%% 
%%%%%%%%%%%%%%%                                            %%%%%%%%%%%%%%%
%%%%%%%%%%%%%%%                                            %%%%%%%%%%%%%%%
\section{Introduction}

The multi-channel Kondo effect\cite{Nozieres} 
has been the subject of intensive 
theoretical
\cite{Andrei,Tsvelik,Cox,Schlottmann,Affleck,Ludwig} 
and experimental\cite{Seaman} studies,
which is characterized by unusual non-Fermi liquid behaviors. 
Its applications are now extended not only to standard dilute 
magnetic alloys, but also to quantum dots, etc. 
Thus far, theoretical and experimental studies 
on the multi-channel Kondo effect have been 
focused on a {\it static} Kondo impurity,
which has been related to the measurements of
the specific heat, the spin susceptibility, the resistivity,
etc. This naturally motivates us to address a 
question  whether such a  nontrivial
phenomenon can be observed in dynamically generated situations.

The photoemission and the inverse photoemission 
may be one of the key experiments 
to study  non-Fermi liquid behaviors,
which reveal the dynamics of a single hole or electron suddenly 
created in the system.  
We here propose 
{\it the dynamically induced multi-channel Kondo effect}, 
when an electron is emitted  from (or added to) 
the Kondo impurity 
by the photoemission (inverse photoemission).
A remarkable point  is that the ground state of 
the system is assumed to be a completely screened Kondo singlet,
and non-Fermi liquid properties are generated by an electron 
or hole suddenly created.
We study low-energy critical properties of the spectrum 
by using the exact solution of
the multi-channel Kondo model \cite{Andrei,Tsvelik} 
combined with boundary conformal field theory (CFT)\cite{bpz,Cardy}. 
We analyze the one-particle Green function for the 
impurity to show its typical non-Fermi liquid behavior. 
It is further demonstrated  that this effect can be observed even in 
a homogeneous system without impurities. To show this explicitly,
we apply the analysis to  the photoemission spectrum 
in a quantum spin chain with spin $S>1/2$. 

This paper is organized as follows. 
In \S 2 we briefly illustrate the idea of 
the dynamically induced multi-channel Kondo effect, and
derive low-energy scaling forms of 
the one-particle Green function. 
We discuss non-Fermi liquid properties in the spectrum
by exactly evaluating the critical exponents. 
In \S 3 the analysis is then  applied to the photoemission 
spectrum for a quantum spin 
chain. Brief summary is given in \S 4.
 We note that preliminary results on this issue have 
been reported in Ref. 11.
%%\cite{fujii3}.

%
%%%%%%%%%%%%%%%                                            %%%%%%%%%%%%%%%
%%%%%%%%%%%%%%%--------------------------------------------%%%%%%%%%%%%%%%
%%%%%%%%%%%%%%%     section 2 DYNAMICAL KONDO EFFECT       %%%%%%%%%%%%%%%
%%%%%%%%%%%%%%%--------------------------------------------%%%%%%%%%%%%%%% 
%%%%%%%%%%%%%%%                                            %%%%%%%%%%%%%%%

\section{Low-Energy Dynamics in the Photoemission Spectrum} 
%%%%%%%%%%%%%%%%%%%%%%%%%%%%%%%%
%%%%%%%%%%\subsection{ scaling forms in the low-energy region}
%%%%%%%%%%%%%%%%%%%%%%%%%%%%%%%%
\subsection{ dynamically induced Kondo effect}

Let us consider the spin-$S$ Kondo impurity which 
is {\it completely screened} by conduction electrons with 
$n(=2S)$ channels. The impurity spin is assumed to be composed 
of $n$ electrons by the strong Hund coupling.
To study the core-electron photoemission spectrum,
we start with spectral properties 
of the impurity Green function,
%%%%%%%%%%%%%%%%%%%%%%%%%%%%%%%
\begin{eqnarray}
   &&
   G(t) = -{\rm i}<{\rm T}{\,}[{\,}d(t){\,}d^{\dagger}(0){\,}]> \cr
   &&{\qquad \ }          
             {
               \displaystyle 
                {\quad}
               \atop
                {
                 \displaystyle 
                  = G^{>}(t) + G^{<}(t),
                }
             }
\label{1}
\end{eqnarray}
%%%%%%%%%%%%%%%%%%%%%%%%%%%%%%%%%%%%%
where $d$ is the annihilation operator for 
one of core electrons which compose the impurity spin 
and T is the conventional time-ordered product. 
Here, $G^{>}(t)$ ($G^{<}(t)$) is the Green function, 
which is restricted in $t>0 \hspace{1mm} (t<0)$.
For the photoemission, we consider $G^{<}(t)$. 
To be specific, we discuss the case that 
a core electron is emitted as depicted in Fig. 1 (a), for which 
the binding energy $-\omega_\alpha $ 
(measured from the Fermi energy) is assumed to be larger than the 
band width $D$. Then in the excited state the overscreening system is
 generated, which is referred to as  
{\it the dynamically induced overscreening Kondo effect}.\cite{fujii3}
At the low-energy regime around $-\omega_\alpha$, 
we may  express the operator as
$d(t) \simeq {\rm e}^{{\rm i}\omega_\alpha t} \phi (t)$ 
where $\phi (t)$ is the corresponding boundary operator\cite{Cardy2,Affleck2} 
in boundary CFT, which characterizes the boundary critical phenomena. 
It is known that the Fermi-edge singularity\cite{Schotte} 
is reformulated by the boundary operator,\cite{Ludwig2} in which
nontrivial effects for the 
overscreening Kondo effect are incorporated in 
$\phi_\alpha (t)$. 
We write down the one-particle Green function 
$G^{<}(t)$ as, 
%%%%%%%%%%%%%%%%%%%%%%%%%%%%%%%%%%%%%%%%%%%%%
\begin{eqnarray}
  {\rm Im}G^{<}(\omega) 
    = \frac{1}{2} \int^{\infty}_{-\infty} 
        <\phi_{\alpha}^{\dagger}(0)\phi_{\alpha}(t)> 
        {\rm e}^{\rm i \omega_\alpha t}
        {\rm e}^{\rm i \omega t} {\rm d} t.
\label{2}
\end{eqnarray}
%%%%%%%%%%%%%%%%%%%%%%%%%%%%%%%%%%%%%%%%%%%%%%%

On the other hand, for the inverse photoemission, 
 an added electron is combined 
with the local spin $S$  
to form higher spin $S+1/2$ by the strong Hund-coupling, 
as shown in Fig. 1 (b). Then we may write
$d(t) \simeq {\rm e}^{-{\rm i}\omega_\beta t} 
 \phi_\beta (t)$,
where $\omega_\beta>D$ is  the energy cost to make
 $S+1/2$ spin, and $\phi_\beta (t)$ 
is another boundary operator which controls 
the undersreening Kondo effect induced  
 by the inverse photoemission. 
We have 
%%%%%%%%%%%%%%%%%%%%%%%%%%%%%%%
\begin{eqnarray}
  {\rm Im}G^{>}(\omega) 
    &=& -\frac{1}{2} \int^{\infty}_{-\infty} 
        <\phi_\beta (t) \phi_\beta^{\dagger}(0)> 
        {\rm e}^{-{\rm i}\omega_\beta t}
        {\rm e}^{\rm i \omega t} {\rm d} t .
\label{2}
\end{eqnarray}
%%%%%%%%%%%%%%%%%%%%%%%%%

In order to  evaluate the critical exponents,
we now employ the idea of finite-size scaling\cite{Cardy} 
in CFT. The scaling form of the correlators
$<\phi_{\alpha}^{\dagger}(0)\phi_{\alpha}(t)>$ and 
$<\phi_{\beta}(t)\phi^{\dagger}_{\beta}(0)>$ are given by 
%%%%%%%%%%%%%%%%%%%%%%%%%%%%%%%%
%%
\begin{eqnarray}
  <\phi_{\alpha}^{\dagger}(0) \phi_{\alpha}(t)>
  &=&
  \sum_{N=0}^{\infty}
   |<0|\phi_{\alpha}^{\dagger}(0)|N>|^2
   {\rm e}^{
    {\rm i }
      \frac{\pi v_F}{l}
      (\Delta_\alpha +N) t} \cr
  &=&
  \left(
  \frac{ \displaystyle \frac{\pi}{2l} }
       { {\rm sinh-} 
          \displaystyle \frac{\pi v_F}{2l} 
           {\rm i} t }
         \right)^{2\Delta_\alpha} 
   \rightarrow
   \frac{1}{ (-{\rm i} v_F t)^{2\Delta_\alpha} } \cr
  & & \cr
 <\phi_{\beta}(t) \phi^{\dagger}_{\beta}(0)>
 &=&
 \sum_{N=0}^{\infty}
   |<N|\phi_{\beta}^{\dagger}(0)|0>|^2
   {\rm e}^{{-\rm i} \frac{\pi v_F}{l}(\Delta_\beta +N) t} \cr
  &=&
   \left(
    \frac
     { \displaystyle \frac{\pi}{2l} }
     { {\rm sinh} 
        { \displaystyle \frac{\pi v_F}{2l} } 
         {\rm i} t }
   \right)^{2\Delta_\beta}
  \rightarrow 
   \frac{1}{ ({\rm i} v_F t)^{2\Delta_\beta} },
\nonumber
\end{eqnarray}
%%%%%%%%%%%%%%%%%%%%%%%%%%%%%%%%%%%%%%%%%%%
in the long-time asymptotic region.
According to the finite-size scaling, 
the boundary dimensions  $\Delta_\alpha$ and $\Delta_\beta$ 
are read from the lowest excitation energy $\Delta E$, 
%%%%%%%%%%%%%%%%%%%%%%%%%%%%%%%%%%%%
\begin{eqnarray}
  \Delta E=\frac{\pi v_{F}}{l} \Delta_\gamma ,
\label{4}
\end{eqnarray}
%%%%%%%%%%%%%%%%%%%%%%%%%%%%%%
with $\gamma=\alpha ,\beta$, where  $l$ corresponds to 
the system size of one dimension in the radial direction.
We thus  end up with the relevant scaling forms as  
%%%%%%%%%%%%%%%%%%%%%%%%%%%%%%%
\begin{eqnarray}
  {\rm Im} G^{<}(\omega)
   &=&
   \frac{\pi}
        {\Gamma(2\Delta_\alpha) {v_F}^{2\Delta_\alpha}}
   \theta(-\omega-\omega_\alpha)
         (-\omega-\omega_\alpha)^{X_{\alpha}}, \cr
   & & \cr
  {\rm Im} G^{>}(\omega)
   &=&
   \frac{-\pi}
        {\Gamma(2\Delta_\beta) {v_F}^{2\Delta_\beta}}
   \theta(\omega-\omega_\beta)
         (\omega-\omega_\beta)^{X_\beta},
\label{5}
\end{eqnarray}
%%%%%%%%%%%%%%%%%%%%%%%%%%%%%%%%%%%%%%%%%%
where $X_\gamma=2\Delta_\gamma -1$ that 
$\gamma$ represents $\alpha$ and $\beta$. 

In both cases, the spectral functions have power-law 
edge singularity due to the {\it dynamically induced
 multi-channel Kondo effect}, which will be shown to
exhibit  non-Fermi liquid properties.

%%%%%%%%%%%%%%%%%%%%%%%%%%%%%%%%%%%%%%%%%%%%%%%%%%%%%%%%
\begin{figure}[htb]
\epsfig{file=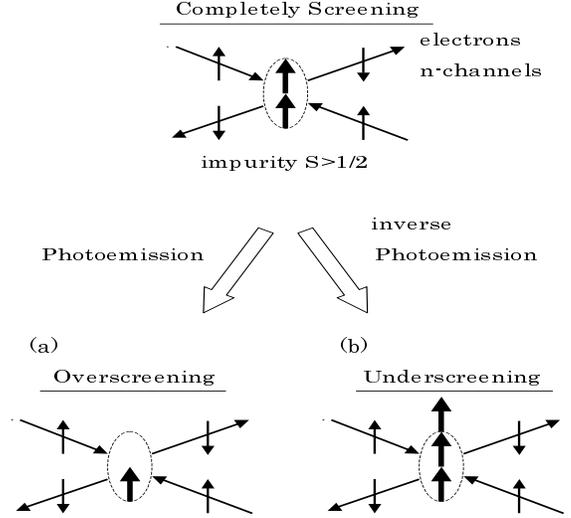,height=90mm,width=90mm}
%%\epsf{file=multi0.eps,height=90mm,width=90mm}
\caption{
Overscreening and underscreening systems dynamically 
generated by the photoemission and the inverse
photoemission, for which 
the impurity spin is assumed to be completely screened 
in the ground state.
}
\label{fig:1}
\end{figure}
%%%%%%%%%%%%%%%%%%%%%%%%%%%%%%%%%%%%%%%%%%%%%%%%%%%%%%%%%%%%%%%%

%%%%%%%%%%%%%%%%%%%%%%%%%%%%%%%%%%%%%%%%%%%%%%%%%%%%%%%
%%
%%\subsection{dynamically induced multi-channel Kondo effect}
\subsection{exact critical properties}
%%%%%%%%%%%%%%%%%%%%%%%%%%%%%%%%%%%%%%%%%%%%%%%%%%%%%%%

We now discuss low-energy critical properties 
by exactly evaluating  $\Delta_\alpha$ and 
$\Delta_\beta$.
To this end, we consider the multi-channel Kondo model, 
%%%%%%%%%%%%%%%%%%%%%%%%%%%
\begin{eqnarray}
{\cal H}
&=&-{\rm i}{\sum_{a,m}}
         \int {\rm d}x
          \psi_{am}^{\dagger}(x) \partial_{x} \psi_{am}(x) \cr
& & \cr       
& &{}+2J {\sum_{a,b,m}}{\sum_{\nu}} \psi_{am}^{\dagger}(0) 
                               \sigma^{\nu}_{ab}
                               \psi_{bm}(0)
                               S^{\nu}
-HS^{z},
\label{6}
\end{eqnarray}
%%%%%%%%%%%%%%%%%%%%%%%%%%%%%%%%%%%%%%%%%%%%%%%%%%%%%
where $\psi^{\dagger}_{am}$ is the creation operator for 
conduction electrons with spin $a=\uparrow,\downarrow$ and 
 orbital indices, $m=1,\cdots,n$.
The exact solution of this model\cite{Andrei,Tsvelik} 
is expressed in terms of the Bethe equations for 
spin rapidities $\lambda_{\alpha}$ 
and charge rapidities $k_{j}$,
%%%%%%%%%%%%%%%%%%%%%%%%%%%%%%%%%%%%%%%%%%%%%%%%%%
\begin{eqnarray}
&&   {\rm e}^{{\rm i} k_{j}L} = \prod_{\alpha=1}^{M}
                             \frac{\lambda_{\alpha}+{\rm i}n/2}
                                  {\lambda_{\alpha}-{\rm i}n/2}\cr
&& \cr
&&
   \frac{\lambda_{\alpha}+1/J+{\rm i}S}
          {\lambda_{\alpha}+1/J-{\rm i}S}
     \left( \frac{\lambda_{\alpha}+{\rm i}n/2}
                 {\lambda_{\alpha}-{\rm i}n/2}
     \right)^{N}
     =
     \prod_{\alpha=1}^{M}
           \frac{\lambda_{\alpha}-\lambda_{\beta}+{\rm i}}
                {\lambda_{\alpha}-\lambda_{\beta}-{\rm i}},
\label{7}
\end{eqnarray}
%%%%%%%%%%%%%%%%%%%%%%%%%%%%%%%%%%%%%%%%%%%%%%%%%%%%%%%
where $N$ is the number of electrons and $L=2l$ is the
one-dimensional system size. 
It is assumed that the impurity with spin $S>1/2 $
is completely screened in the ground state. Then, 
the core-level photoemission suddenly reduces 
the impurity spin, thus inducing 
the overscreening Kondo effect with $n-2S=1$. 
As for the underscreening effect induced by 
the inverse photoemission, the condition 
is replaced by $n-2S=-1$. At zero temperature 
the ground-state properties are described by the $n-$th 
order string solutions,\cite{Andrei,Tsvelik}  
%%%%%%%%%%%%%%%%%%%%
%%
\begin{eqnarray}
  \lambda_l^{n,\alpha}=\lambda_l^n +\frac{\rm i}{2}
                        (n+1-2\alpha),
\label{8}
\end{eqnarray}
%%%%%%%%%%%%%%%%%%%%
%%
where $\alpha=1, \cdots ,n$ and 
$l=1, \cdots ,M_n$ which is restricted by $M=nM_n$. 
Here, $n$ represents the number of orbitals.
It is well known that 
a naive application of finite-size techniques based on the 
string hypothesis turns out to fail 
for the overscreening case at zero magnetic 
field.\cite{Reshetikhin,Klumper,Kuniba,Fujimoto}
This difficulty comes from an improper treatment of
 the $\rm Z_n$ symmetry sector in terms of the string solutions.
However, as long as the finite magnetic field is concerned, 
we can use the Bethe equations to describe  its 
critical properties. 
We will separately discuss the case of zero-magnetic field, 
by incorporating $\rm Z_n$ 
sector correctly.\cite{Reshetikhin,Klumper,Kuniba,Fujimoto}
By applying standard procedures to eq.(\ref{7}), 
it is straightforward to exactly
 evaluate the lowest excitation energy in magnetic fields,
for which one of the impurity electrons is assumed to be 
removed from the system, 
%%%%%%%%%%%%%%%%%%%
\begin{eqnarray}
  \Delta E=\frac{\pi v_{F}}{l} \cdot 
            \left(
             \frac{\delta_\alpha^2}{4n}+ 
             n(n_{{\rm imp}})^2
            \right).
\label{9}
\end{eqnarray}
%%%%%%%%%%%%%%%%%%%
Although the above finite-size correction is 
apparently similar to that for 1D solvable systems 
with a static impurity or boundaries,\cite{e1,e2,e3,e4,e5,e6,e7}
 {\it the final-state interaction} 
induced by photoemission is included  in 
the present case.\cite{fujii} Thus
all the features which are governed by the dynamical Kondo effect
can be read from this quantity. 

By applying the finite-size scaling
in eq. (\ref{4}), the critical exponent 
$X_{\gamma}$ in eq. (\ref{5}) is now obtained as, 
%%%%%%%%%%%%%%%%%%%%
%%
%%\begin{eqnarray}
%%  \Delta_\gamma=\frac{\delta_\gamma^2}{4n}+ n(n_{{\rm imp}})^2,
%%\label{10}
%%\end{eqnarray}
%%%%%%%%%%%%%
%%%%%%%%%%%%%%%%%%%%%%%%%%%%%%%%%%%%%%
\begin{eqnarray}
 X_{\alpha}=\frac{\delta_{\alpha}^{2}}{2n}+2n(n_{{\rm imp}})^2 -1,
\label{11}
\end{eqnarray}
%%%%%%%%%%%%%%%%%%%%%%%%%%%%%%%%%%%%%%
where $\delta_\alpha$ is the charge scattering phase shift,
which is caused by a created hole
as in  the ordinary Fermi edge singularity.\cite{Schotte} 
This term depends on the detail of potential scattering. 
It is mentioned that the second term with  the phase shift $n_{\rm imp}$ 
is caused by the Kondo effect, which is explicitly evaluated as,
%%%%%%%%%%%%%%%%%%%%%%%%%%%%%%%%%%%%%%%5%%
\begin{eqnarray}
 n_{\rm imp} = \int_{-\infty}^{\lambda_0}
         \sigma_{\rm imp} (\lambda) {\rm d}\lambda ,
\label{12}
\end{eqnarray}
%%%%%%%%%%%%%%%%%%%%%%%%%%%%%%%%%%%%%%%%%%%%
where
%%%%%%%%%%%%%%%%%%%%%%%%%%%%%%%%%%%%%%%%%%%%%
\begin{eqnarray}
 \sigma_{\rm imp} (\lambda) 
  &=& 
   \sigma^0_{\rm imp} (\lambda +1/J) \cr
  & & \cr
  & & 
        {} -\int_{-\infty}^{\lambda_0}
        G_n(\lambda -\lambda^{'})
        \sigma_{\rm imp} (\lambda^{'}) 
        {\rm d}\lambda^{'}. 
\label{13}
\end{eqnarray}
%%%%%%%%%%%%%%%%%%%%%%%%%%%%%%%%%%%%%%
Here $\sigma^0_{\rm imp} (\lambda)$ and 
$G_n(\lambda)$ are determined by 
%%%%%%%%%%%%%%%%%%%%%%%%%%%%%%%%%%%%%%%%%%%
\begin{eqnarray}
&& \sigma^0_{\rm imp} (\lambda)
    = \frac{1}{\pi}
       \sum^{{\rm min}(n,2S)}_{l=1} 
        \frac{ {\displaystyle \frac{1}{2}} (n+2S+1-2l)}
         {\lambda^2+
          { \displaystyle \frac{1}{4} }(n+2S+1-2l)^2}, \cr
&& \ \cr
&& G_n (\lambda)
    =\frac{1}{\pi} 
      \frac{n}{\lambda^2 +n^2}
     + \frac{2}{\pi} 
        \sum^{n-1}_{\alpha =1}
         \frac{ {\displaystyle \frac{1}{2} }(2n-2\alpha)}
         {\lambda^2 +
           { \displaystyle \frac{1}{4} }(2n-2\alpha)^2}.
\label{14}
\end{eqnarray}
%%%%%%%%%%%%%%%%%%%%%%%%%%%%%%%%%%%%%%
The key quantity, $n_{\rm imp}$, is obtained by 
using Wiener-Hopf method, \cite{Andrei,Tsvelik} 
%%%%%%%
\begin{eqnarray}
  && 
     n_{{\rm imp}}= \frac{S}{n}-
                    \left(
                     \frac{S}{n}-\frac{1}{2}
                    \right)
                   \theta 
                    \left(
                     \frac{S}{n}-\frac{1}{2}
                    \right)  \cr
  && \ \cr
  && \ \quad 
                  +\frac{{\rm i}}{4\pi^{\frac{3}{2}}}
                    \int_{-\infty}^{\infty}
                    \frac{{\rm d}\omega}{\omega-{\rm i}0}
                   {\rm e}^{-{\rm i}2\omega 
                   \log{\frac{H}{T_{H}}}}
                   \frac{\Gamma \left(1+{\rm i}\omega \right)
                         \Gamma \left(
                                      1/2-{\rm i}\omega
                                \right)}
                        {\Gamma(1+{\rm i} n\omega)} \cr
  && \ \cr
  && \quad \qquad \qquad \!
                  \cdot
                  \left(
                   \frac{{\rm i}n \omega +0}{\rm e}
                  \right)^{{\rm i}n \omega}
                  \frac{
                   {\rm e}^{-\pi|n-2S| |\omega| }
                  -{\rm e}^{-\pi(n+2S)|\omega| }
                       }
                       {
                        1-{\rm e}^{-2\pi n|\omega|}
                       }.
\nonumber
\end{eqnarray}
%%%%%%%%%%%%%%%%%%%%%%%%%

%%%%%%%%%%%%%%%%% FIGURE %%%%%%%%%%%%%%%%%%%%%%%%%%%%%%%%%%%%%%%
\begin{figure}[htb]
\epsfig{file=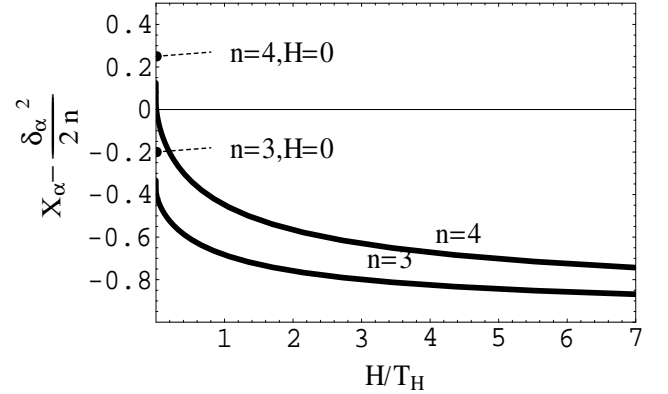,height=56mm,width=84mm}
\caption{
The critical exponent $X_{\alpha}$ 
as a function of magnetic fields 
for $n=3$ and 4.  For clarity, we have subtracted 
the contribution from the ordinary
charge scattering,  $\delta_{\alpha}^{2}/2n$.
The energy unit is the
magnetic Kondo temperature $T_{H}$,\protect{\cite{Andrei,Tsvelik}} 
which is expressed in terms of the ordinary Kondo temperature
$T_{K}$ via
$\displaystyle
T_{H}=[2\pi(n/2{\rm e})^{n/2}/\Gamma (n/2)] T_{K}$. 
At  zero-magnetic field, the exponents 
show a discontinuity, which 
is related to the $\rm Z_n$ symmetric sector, as
mentioned in the text.
\protect{\cite{Zamolodchikov,Tsvelik2,Fujimoto}}
}
\label{fig:2}
\end{figure}
%%%%%%%%%%%%%%%%%%%%%%%%%%%%%%%%%%%%%%%%%%%%%%%%%%%%%%%

In Fig. 2 we display the critical exponent $X_{\alpha}$ 
for the overscreening case as a  function of the  magnetic field.
Note that the magnetic-field dependence of
$X_\alpha$ without $\delta_{\alpha}^2 /2n$  
is determined by the dynamical Kondo effect,
because the charge scattering phase shift 
$\delta_\alpha$ does not depend on magnetic fields. 
Particularly in weak magnetic field, $H<<T_H$, 
the obtained exponent behaves as 
%%%%%%%%%%%%%%%%%%%%%%%%%%%%%%%%%%%%%%
\begin{eqnarray}
 X_{\alpha} \rightarrow 
   \frac{\delta^{2}}{2n}+
    \left( \frac{S}{n}-
     {\rm const} \cdot 
      \left( 
       \frac{H}{T_{H}}   
      \right)^{\textstyle \frac{2}{n}}
    \right)^2 -1.
\label{15}
\end{eqnarray}
%%
%%%%%%%%%%%%%%%%%%%%%%%%%%%%%%%%%%%%%%%%%%%%%%%
It is seen  that the phase shift, 
$n_{\rm imp}$, gives rise to the 
anomalous magnetic-field dependence of the exponent. 
This non-Fermi liquid behavior 
is characteristic of the overscreening effect.\cite{Andrei,Tsvelik}
Another interesting feature in the overscreening effect 
appears at $H=0$. 
We recall here that the symmetry is enhanced from U(1) to
SU(2) at $H=0$, for which  the boundary dimension  $\Delta_s$
for the spin sector is analytically obtained by
employing fusion rules hypothesis proposed by Affleck and Ludwig,
\cite{Affleck3,Affleck}
%%%%%%%%%%%%%%%%%%
\begin{eqnarray}
 \Delta_s =\frac{S(S+1)}{n+2},
\label{16}
\end{eqnarray}
%%%%%%%%%%%%%%%%%%
which is a typical conformal dimension 
for level-$n$ SU(2) Kac-Moody algebra. 
%%a remarkable point is that 
%%the quantum number 
%%unusually represents the impurity spin $S$. 
Note that the critical exponent shows a discontinuity
at $H=0$,  
%%%%%%%%%%%%%%%%%%%%%%%%%%%%%%%%
\begin{eqnarray}
 X_\alpha (H=0)-X_\alpha (H\rightarrow 0)
 =2 \cdot \frac{S(S+1)}{n+2}-2 \cdot \frac{S^2}{n},
\label{}
\end{eqnarray}
%%%%%%%%%%%%%%%%%%%%%%%%%%%%%
which is caused by the fact that 
$\rm Z_n$ symmetric sector is massless only 
at $H=0$,\cite{Zamolodchikov,Tsvelik2}
as already mentioned.

We now move to the underscreening case 
induced by the inverse photoemission.
The calculated critical exponent $X_{\beta}$ 
is shown as a function of magnetic fields
in Fig. 3.
For weak magnetic field, 
the obtained exponent $X_\beta$ behaves as 
%%%%%%%%%%%%%%%%%%%%%%%%%%%%%%%%%%%%%%%%%%%%
\begin{eqnarray}
 X_{\beta} \rightarrow \frac{\delta_{\beta}^{2}}{2n}+
                       n \left( \frac{1}{2}-
                                \frac{1}{\log (H/T_H)}
                         \right)^2 -1,
\label{18}
\end{eqnarray}
%%%%%%%%%%%%%%%%%%%%%%%%%%%%%%%%%%%%%%%%%%%%%%
which is characteristic of the underscreening system. 
In contrast to the overscreening case, there is 
no discontinuity in the exponent in this case,
$X_\beta (H=0)= X_\beta (H \rightarrow 0)$. 

This completes a general description of the dynamically induced
Kondo effect.  An important point to be emphasized 
 is that this kind of phenomenon
may be observed not only for impurity systems but also for other
related quantum systems, which will be explicitly 
discussed  in the next section.

%%%%%%%%%%%%%%%%%%%%%%%%%%%%%%%%%%%%%%%%%%%%%%%%%%%%%%%%
\begin{figure}[htb]
\epsfig{file=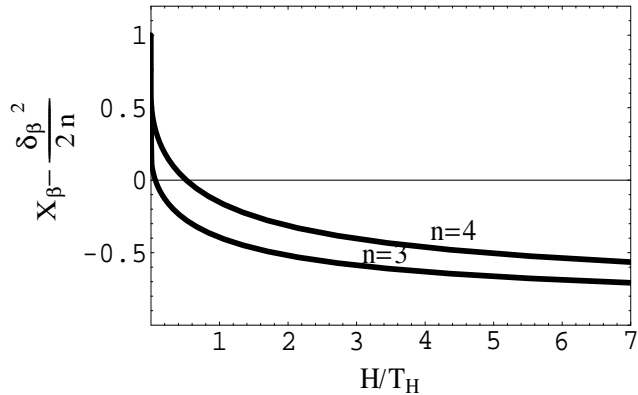,height=56mm,width=84mm}
\caption{
The critical exponent $X_{\beta}$ for the underscreening 
Kondo effect, induced by the inverse photoemission,
as a function of magnetic fields  for $n=3$ and 4. 
The charge phase shift $\delta_{\beta}^{2}/2n$ is subtracted
for simplicity, as before.
}
\label{fig:3}
\end{figure}
%%%%%%%%%%%%%%%%%%%%%%%%%%%%%%%%%%%%%%%%%%%%%%%%%%%%%%%

%%%%%%%%%%%%%%%                                            %%%%%%%%%%%%%%%
%%%%%%%%%%%%%%%                                            %%%%%%%%%%%%%%%
%%%%%%%%%%%%%%%--------------------------------------------%%%%%%%%%%%%%%%
%%%%%%%%%%%%%%%   section 3 APPLICATION FOR SPIN CHAINS    %%%%%%%%%%%%%%%
%%%%%%%%%%%%%%%--------------------------------------------%%%%%%%%%%%%%%% 
%%%%%%%%%%%%%%%                                            %%%%%%%%%%%%%%%
%%%%%%%%%%%%%%%                                            %%%%%%%%%%%%%%%
%
\section{Application to Spin Chains}

We now wish to demonstrate that the dynamically induced  Kondo effect
proposed here may be observed for gapless quantum spin systems 
{\it which do not possess impurities}. As 
an example, we consider an integrable antiferromagnetic spin 
chain with spin $S>1/2$, for which the exact solution  is
available even for the case with doped holes.\cite{Frahm}
The photoemission suddenly removes one electron from the spin
system, and thus bears an {\it impurity site} with spin $S-1/2$.
In the final state, this impurity spin is screened 
by host spins, and as a result the overscreening 
Kondo effect may be dynamically induced. It is remarkable 
that the induced impurity in this case
can move through the lattice via the exchange interaction, and 
the edge singularity is thus governed by 
a {\it mobile} multichannel  Kondo impurity.\cite{tsukamoto} 

Let us consider the gapless $S=1$ spin chain with a mobile
$S=1/2$ impurity as an example. Although in more general situations
including  non-integrable models, 
the higher spin should be a half-odd integer,
the present treatment can be straightforwardly 
extended to such cases.
We here consider an integrable spin chain derived by the 
quantum inverse scattering method, \cite{Frahm}
%%%%%%%%%%%%%%%%%%%%%%%%%%%%%%%%%%%
\begin{eqnarray}
 &{\cal H}&=\sum_{i=1}^{L}
         -(1-\delta_{S_{i}S_{i+1},1}) {\cal P}_{i,i+1} 
          ({\bf S}_{i} \cdot {\bf S}_{i+1}) \cr
 & & \ \cr
 &+& \frac{1}{2} \left( \frac{1}{S_{i}S_{i+1}}
                           {\bf S}_{i} \cdot {\bf S}_{i+1} -1
                           +\delta_{S_{i}S_{i+1},1}
                            [1-({\bf S}_{i} \cdot {\bf S}_{i+1})^{2}] 
                         \right) 
\nonumber
\end{eqnarray}
%%%%%%%%%%%%%%%%%%%%%%%%%%%%%%%%%%%%%%%%%
where the spin ${\bf S_i}^2=S_i(S_i+1)$ with $S_i=1$ or 1/2,
and ${\cal P}_{ij}$ permutes the spin on sites $i$ and $j$.
%%Note that we will deal with the case that there is only 
%%one $S=1/2$ mobile impurity created by the photoemission.

In order to deal with the excited states when an electron 
is emitted from the spin chain, 
we write down the  Bethe equations for the spin-$S$ chain
with {\it one hole} being doped, 
%%%%%%%%%%%%%%%%%%%%%%%%%%%%%%%%%%%%%%%%%%%5
\begin{eqnarray}
 \left(
  \frac{\lambda_j+{\rm i}S}{\lambda_j-{\rm i}S}
 \right)^L
 &=&
 \frac{\lambda_j -\nu_ -{\rm i}/2}
      {\lambda_j -\nu_ +{\rm i}/2}
 \prod^{N_{\downarrow}+1}_{k \neq j}
 \frac{\lambda_j -\lambda_k +{\rm i}}
      {\lambda_j -\lambda_k -{\rm i}}, \cr
  & & \ \cr
1 &=&
  \prod^{N_{\downarrow}+1}_{k=1}
  \frac{\nu -\lambda_k +{\rm i}/2}
       {\nu -\lambda_k -{\rm i}/2},
\label{19}
\end{eqnarray}
%%%%%%%%%%%%%%%%%%%%%%%%%%%%%%%%%%%%%%%%%%%%%%%%%%
where $N_{\downarrow}$ is the number of down spins and
$\lambda_j (j=1,\cdots ,N_{\downarrow}+1)$ are spin rapidities. 
Note that the hole rapidity $\nu$ appears in the 
above equation, which characterizes a massive charge excitation 
suddenly created. 
Thus, $\nu$ specifies the hole momentum $q$. 
It is seen  that the above spin-charge scattering term, 
which describes the final-state interaction,
corresponds to the impurity term in eq. (\ref{7}). 

The manipulation illustrated in the previous section 
enables us to exactly  calculate the scaling dimension 
for the one-particle Green function via
the finite-size corrections,\cite{Sorella,fujii}
%%%%%%%%%%%%%%%%%%%%%%%%%%%%%%%%%
\begin{eqnarray}
 x(\nu)=\frac{1}{4\xi_{2S}^2}(1-n_{\rm imp}(\nu))+
        \xi_{2S}^2(\frac{1}{2}-d_{\rm imp}(\nu))^2.
\label{20}
\end{eqnarray}
%%%%%%%%%%%%%%%%%%%%%%%%%%%
The quantity $\xi_{2S} \equiv \xi_{2S}(\lambda_0)$ 
often referred to as the dressed charge
\cite{dressed,fk,kawakami}
 is given by 
%%%%%%%%%%%%%%%%%%%%%%%%%%%%%
\begin{eqnarray}
 \xi_{2S} (\lambda)=1-\int^{\lambda_0}_{-\lambda_0} 
               K_{2S}(\lambda -\lambda^{'})
               \xi_{2S} (\lambda^{'}),
\label{21}
\end{eqnarray}
%%%%%%%%%%%%%%%%%%%%%%%%%%%%%%%%
where 
%%%%%%%%%%%%%%%%%%%%%%%%%%%%%%%%
\begin{eqnarray}
 K_{2S}(\lambda)&=&\frac{1}{\pi}
               \frac{\frac{1}{2}(4S)}
                    {\lambda^2+\frac{1}{4}(4S)^2} \cr
            & &{\quad}+
               \frac{2}{\pi}
               \sum_{l=1}^{2S+1}
               \frac{\frac{1}{2}(4S-2l)}
                    {\lambda^2+\frac{1}{4}(4S-2l)^2}, 
\label{22}
\end{eqnarray}
%%%%%%%%%%%%%%%%%%%%%%%%%%%%%%%%%%%%%%%%%%%
and the cut-off parameter  $\lambda_0$ is related to 
the magnetization $m$, 
%%%%%%%%%%%%%%%%%%%%%%%%%%%%%%
\begin{eqnarray}
    S-m=2S \int^{\lambda_0}_{-\lambda_0}
              \rho_{2S} (\lambda) {\rm d}\lambda. 
\label{23}
\end{eqnarray}
%%%%%%%%%%%%%%%%%%%%%%%%%%%%%
The density function $\rho_{2S}$ is determined by the following 
integral equation, 
%%%%%%%%%%%%%%%%%%%%%%%%%%%%%%%%%%%%%%
\begin{eqnarray}
 \rho_{2S}(\lambda)=\frac{1}{2\pi}
                  \Theta_{2S,2S}^{'}(\lambda)
                  -\int^{\lambda_0}_{-\lambda_0} 
                    K_{2S}(\lambda -\lambda^{'})
                    \rho_{2S} (\lambda^{'}). 
\label{24}
\end{eqnarray}
%%%%%%%%%%%%%%%%%%%%%%%%%%%%%%%%%%%%%%%%%%%%%%%%
%%The dressed charge features the U(1) critical line 
%%%of $c=1$ CFT. 
We stress that two key quantities $n_{\rm imp}(\nu)$ and 
$d_{\rm imp}(\nu)$, which contain the effect of a mobile
impurity, are introduced in eq. (\ref{20}), 
%%%%%%%%%%%%%%%%%%%%%%%%%%%%%%
\begin{eqnarray}
 n_{\rm imp}(\nu)&=&\int^{\lambda_0}_{-\lambda_0} 
               \rho_{\rm imp} (\lambda) {\rm d} \lambda \cr
             & & \ \cr
 d_{\rm imp}(\nu)&=&-\frac{1}{2}
                \left(
                 \int^{\infty}_{\lambda_0}-
                 \int^{-\lambda_0}_{-\infty}
                \right)
               \rho_{\rm imp} (\lambda) {\rm d} \lambda
\label{25}
\end{eqnarray}
%%%%%%%%%%%%%%%%%%%%%%%%%%%%%%
where 
%%%%%%%%%%%%%%%%%%%%%%%%%%%%%%%%%%%%%%%%%%%
\begin{eqnarray}
 \rho_{\rm imp} (\lambda)=\frac{1}{2\pi}
                  \Theta_{2S,1}^{'}(\lambda -\nu)
                  -\int^{\lambda_0}_{-\lambda_0} 
                    K_{2S}(\lambda -\lambda^{'})
                    \rho_{\rm imp} (\lambda^{'}).
\nonumber
\end{eqnarray}
%%%%%%%%%%%%%%%%%%%%%%%%%%%%%%%%%%
Here we have introduced the phase function, 
%%%%%%%%%%%%%%%%%%%%%%%%%%%%%%
\begin{eqnarray}
 \frac{1}{2\pi} \Theta_{n,k}^{'}(\lambda)
 =
 \frac{1}{\pi}
  \sum_{l=1}^{{\rm min}(n,k)}
               \frac{\displaystyle \frac{1}{2}(n+k+1-2l)}
                    {\lambda^2+
                     \displaystyle \frac{1}{4}(n+k+1-2l)^2}.
\label{26}
\end{eqnarray}
%%%%%%%%%%%%%%%%%%%%%%%%%%%%%%%%%%%%%%
These quantities are alternatively represented in 
terms of the phase shifts $\delta_L$ and $\delta_R$ 
at the left and right Fermi points in massless spin
excitations: 
$n_{\rm imp}(\nu)=(\delta_L +\delta_R)/2\pi, 
d_{\rm imp}(\nu)=(\delta_L -\delta_R)/2\pi$. 
We mention that the asymmetric phase shift $d_{\rm imp}(\nu)$ 
is inherent in a  {\it mobile} Kondo impurity, 
different from a localized impurity in \S 2.  
Note that the scaling dimension $x$ depends on 
the hole momentum $q$ through the asymmetric phase shift. 

Let us now discuss low-energy critical properties 
in the photoemission spectra. We write down 
the one-particle Green function 
which depends on the momentum $q$, 
%%%%%%%%%%%%%%%%%%%%%%%%%%%%%%%%%%%%%%
\begin{eqnarray}
  {\rm Im}G(q,\omega) \propto 
                     (\omega-\omega_c(q))^{X(q)},
\label{27}
\end{eqnarray}
%%%%%%%%%%%%%%%%%%%%%%%%%%%%%%
with $X(q)=2x-1$, where $x$ is the scaling dimension 
in eq. (\ref{20}) 
and $\omega_c(q)$ is the dispersion of the charge 
excitation generated  by the photoemission. 
In Fig. 4 we show the obtained critical exponent 
as a function of the momentum $q$ for the $S=1$ case.

%%%%%%%%%%%%%%%%%  FIGURE %%%%%%%%%%%%%%%%%%%%%%%%%%%%%%%
\begin{figure}[htb]
\epsfig{file=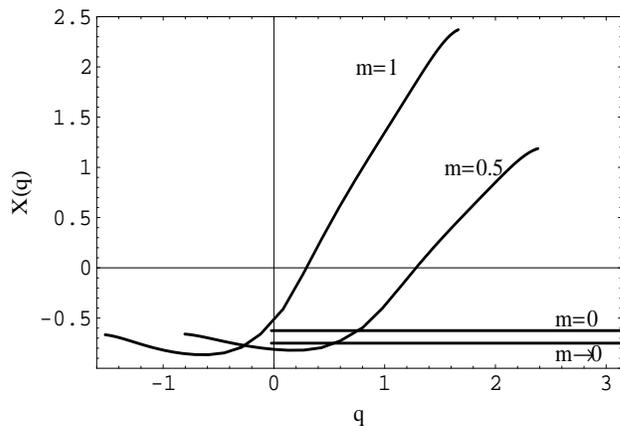,height=56mm,width=84mm}
\caption{
The momentum-dependent exponent $X_{\alpha}(q)$ 
for an integrable (massless) $S=1$ quantum spin chain,
for which 
the overscreening Kondo effect with the orbital degeneracy $n=2$
is induced by the photoemission.
The quantity $m$ is the magnetization of the spin chain in magnetic
fields.
}
\label{fig:4}
\end{figure}
%%%%%%%%%%%%%%%%%%%%%%%%%%%%%%%%%%%%%%%%%%%%%%%%%%%%%%%

This anomalous power-law behavior 
and the momentum dependence of $X(q)$ are caused 
by a suddenly induced {\it mobile} Kondo impurity. 
As we discussed in \S 2,   
the discontinuity of the exponent $X$ at $m=0$ is caused by 
the $\rm Z_n$ symmetric sector.\cite{Zamolodchikov,Alcaratz}. 

In this way, the dynamically induced Kondo effect proposed here
may be expected to be observed in photoemission 
experiments for quantum spin chains. 
In order for this phenomenon to be observed, there may be
several problems to be resolved; preparation of  a 
proper 1D sample, experiments with high resolution, etc.
Anyway, it is desired to experimentally find or
synthesize rather ideal spin chain systems
without the magnetic order even at low temperatures.

%%%%%%%%%%%%%%%                                            %%%%%%%%%%%%%%%
%%%%%%%%%%%%%%%                                            %%%%%%%%%%%%%%%
%%%%%%%%%%%%%%%--------------------------------------------%%%%%%%%%%%%%%%
%%%%%%%%%%%%%%%        section 4 SUMMARY                   %%%%%%%%%%%%%%%
%%%%%%%%%%%%%%%--------------------------------------------%%%%%%%%%%%%%%% 
%%%%%%%%%%%%%%%                                            %%%%%%%%%%%%%%%
%%%%%%%%%%%%%%%                                            %%%%%%%%%%%%%%%
%
\section{summary}

In summary we have proposed the  multi-channel Kondo effect
dynamically induced by the photoemission and the inverse photoemission,
for which the ground state is a completely screened Kondo singlet.
By studying low-energy critical properties in the photoemission spectra,
it has been found that the anomalous behavior generated by this effect
is indeed characteristic of the multi-channel Kondo
system. In particular, we have demonstrated that the idea proposed here can
be directly applied to homogeneous quantum spin systems without any impurity.
It has been shown in this case that a {\it mobile} Kondo impurity 
suddenly created by the photoemission gives rise to 
the momentum-dependent anomalous exponent. 

Although we have mainly focused on the photoemission spectra 
for magnetic impurity systems in this paper,  
it should be noted that the idea can be 
directly applied to the photoemission spectrum 
in quantum dot systems, for which a multilevel quantum dot
with the Hund coupling plays a role of the Kondo impurity. 
In this connection, it is also 
interesting to apply a similar idea to 
the optical absorption spectra in a quantum dot, as recently demonstrated.
\cite{kikoin,fujii2} In such cases, not only the linear but also 
the non-linear optics play an important role,\cite{fujii2} which may
provide interesting phenomena related to the dynamically
induced Kondo effect. 

%%%%%%%%%%%%%%%%%%%%%%%%%%%%%%%%%%
\section*{Acknowledgements}
%%%%%%%%%%%%%%%%%%%%%%%%%%%%%%%%%%%%%
This work was partly supported by a Grant-in-Aid from the Ministry
of Education, Science, Sports and Culture, Japan.
%%%%%%%%%%%%%%%%%%%%%%%%%%%%%%%%%%%%%%%%%%%%%%%

%

\begin{thebibliography}{99}
%
\bibitem{Nozieres}
P. Nozi$\grave{\rm e}$res and Blandin, 
J. Phys. (Paris) {\bf 41}, 193 (1980).
%
\bibitem{Andrei}
N. Andrei and C. Destri, 
Phys. Rev. Lett. {\bf 52}, 364 (1984), 
%%N. Andrei, N. Furuya and J. H. Lowenstein, 
%%{\it Rev. Mod. Phys.} {\bf 55}}, 331 (1983). 
%
\bibitem{Tsvelik}
A. M. Tsvelik and P. B. Wiegmann, 
Z. Phys. B {\bf 54}, 201 (1984).
%
\bibitem{Cox}
D. L. Cox, 
Phys. Rev. Lett. {\bf 59}, 1240 (1991); 
D. L. Cox and A. Zawadowski, 
{\it Exotic Kondo Effects in Metals}, 
(Taylor and Francis, 1999). 
%
\bibitem{Schlottmann}
P. Schlottmann and P. D. Sacramento, 
{\it Adv. Phys.} {\bf 42}, 641 (1993).
%
\bibitem{Affleck}
I. Affleck and A. W. W. Ludwig, 
Nucl. Phys. B {\bf 360}, 641 (1991).
% 
\bibitem{Ludwig}
A. W. W. Ludwig and I. Affleck, 
Phys. Rev. Lett. {\bf 67}, 3160 (1991); 
I. Affleck and A. W. W. Ludwig,
Phys. Rev. B {\bf 48}, 7397 (1993).
%
\bibitem{Seaman}
C. L. Seaman et.al., 
Phys. Rev. Lett. {\bf 67}, 2882 (1991).
%
%
%%%%%%%%%%%%%%%%%%%%%%%%%%%%%%%%%%%%%%%%%%%%%%%%%%%%%%%%%%%%%%%%%%
\bibitem{bpz} 
A. A. Belavin, A. M. Polyakov and A. B. Zamolodchikov, 
Nucl. Phys. B {\bf 241}, 333 (1984).
%
\bibitem{Cardy} 
J. L. Cardy, 
Nucl. Phys. B {\bf 270}, 186 (1986).
% 
%%%%%%%%%%%%%%%%%%%%%%%%%%%%%%%%%%%%%%%%%%
% 
\bibitem{fujii3}
T. Fujii and N. Kawakami,  
Physica B {\bf 281}, 406 (2000).
%%
\bibitem{Cardy2} 
J. L. Cardy, 
Nucl. Phys. B {\bf 324}, 581 (1989).
% 
%%%%%%%%%%%%%%%%%%%%%%%%%%%%
\bibitem{Affleck2}
I. Affleck and A. W. W. Ludwig,
Phys. Rev. Lett. {\bf 67}, 161 (1991).
%
%
\bibitem{Schotte}
K. D. Schotte and U. Schotte, 
Phys. Rev. {\bf 182}, 479 (1969).
%
\bibitem{Ludwig2}
I. Affleck and A. W. W. Ludwig, 
J. Phys. A: Math. Gen. {\bf 27}, 5375 (1994).
%%%%%%%%%%%%%%%%%%%%%%%%%%%%%%%%%%%%%%%%%%
\bibitem{Reshetikhin}
N. Yu. Reshetikhin, 
Lett. Math. Phys. {\bf 7}, 205 (1983).
%
%
\bibitem{Klumper}
A. Klumper and P. A. Pearce, 
Physica A {\bf 183}, 304 (1992).
%
%
\bibitem{Kuniba}
A. Kuniba, T. Nakanishi and J. Suzuki, 
Int. J. Mod. Phys. A {\bf 9}, 5215 (1994); 
{\bf 9}, 5267 (1994).
%
%
\bibitem{Fujimoto}
S. Fujimoto and N. Kawakami, 
Phys. Rev. B {\bf 52}, R10132 (1995).
%
%%%%%%%%%%%%%%%%%%%%%%%%%%%%%%%%%%%%%%%%%%%%%%
%
%\bibitem{woy}
%F. Woynarovich: 
%J. Phys. A {\bf 22} (1989) 4243.
%
%%%%%%%%%%%%%%%%%%%%%%%%%%%%%%%%%%%%%
\bibitem{e1}
H. Schulz, 
J. Phys. C {\bf 18}, 581 (1985).
\bibitem{e2}
H. Asakawa and M. Suzuki, 
J. Phys. A {\bf 29}, 225 (1996); 
A {\bf 30}, 3741 (1997).
\bibitem{e3}
A. A. Zvyagin and P. Schlottmann, 
Phys. Rev. B {\bf 54}, 15191 (1996).
\bibitem{e4}
S. Fujimoto and N. Kawakami, 
Phys. Rev. B {\bf 54}, 5784 (1996).
\bibitem{e5}
G. Bed\"urftig, F. Essler and H. Frahm, 
Phys. Rev. Lett. {\bf 77}, 5098 (1996);
Nucl. Phys. B {\bf 489}, 697 (1997).
\bibitem{e6}
H. Shiroishi and M. Wadati, 
J. Phys. Soc. Jpn. {\bf 66}, 1 (1997).
\bibitem{e7}
T. Deguchi and R. Ye, 
J. Phys. A {\bf 30} 8129 (1997). 
%%%%%%%%%%%%%%%%%%%%%%%%%%%%%%%%%%%%%%%%%%%%%%%%%%%%%%%%%%%%%
\bibitem{fujii} 
T. Fujii, Y. Tsukamoto and N. Kawakami, 
J. Phys. Soc. Jpn. {\bf 66}, 2552 (1997); 
{\bf 68}, 151 (1999).
%%%%%%%%%%%%%%%%%%%%%%%%%%%%%%%%%%%%%%%%%%%%%%%%%%%%%%%%%%%%%
\bibitem{Zamolodchikov}
A. B, Zamolodchikov and V. A. Fateev, 
Sov. Phys. JETP {\bf 62}, 215 (1985).
%
\bibitem{Tsvelik2}
A. M. Tsvelik, 
Nucl. Phys. B {\bf 305} [FS23], 675 (1988).
%
%%%%%%%%%%%%%%%%%%%%%%%%%%%%%%%%%%%%%%%%%%%%%%%%%%%%%%%%%%%%%
%
\bibitem{Affleck3}
I. Affleck and A. W. W. Ludwig, 
Nucl. Phys. B {\bf 352}, 849 (1991).
% 
%%%%%%%%%%%%%%%%%%%%%%%%%%%%%%%%%%%%%%%%%%%%%%%%%%%%%%%%%%%%%
\bibitem{Frahm}
H. Frahm, M. P. Pfannm$\ddot{\rm u}$ller and A. M. Tsvelik, 
Phys. Rev. Lett. {\bf 81}, 2116 (1998); 
H. Frahm, 
Nucl. Phys. B {\bf 559}, 613 (1999). 
%
\bibitem{tsukamoto} 
Y. Tsukamoto, T. Fujii and N. Kawakami, 
Phys. Rev. B {\bf 58}, 3633 (1998); 
Eur. Phys. J. B {\bf 5}, 479 (1998).
%%%%%%%%%%%%%%%%%%%%%%%%%%%%%%%%%%%%%%%%%%%%%%%%%%%%%%%%%%%%%
\bibitem{Alcaratz}
F. C. Alcaratz and M. J. Martins, 
J. Phys. A: Math. Gen. {\bf 22}, 1829 (1989).
%
%%%%%%%%%%%%%%%%%%%%%%%%%%%%%%%%%%%%%%%%%%%%%%%%%%%%%%%%%%%%%
\bibitem{Sorella}
S. Sorella and A. Parola, 
Phys. Rev. Lett. {\bf 76}, 4604 (1996); 
Phys. Rev. B {\bf 57}, 6444 (1998)
%%%%%%%%%%%%%%%%%%%%%%%%%%%%%%%%%%%%%%%%%%%%%%%%%%%%%%%%%%%%%
\bibitem{dressed}
N. M. Bogoliubov, A. G. Izergin and  V. E. Korepin,
Nucl. Phys. B {\bf 275}, 687 (1986). 
%
\bibitem{fk} 
H. Frahm and V. Korepin: 
Phys. Rev. B {\bf 42} (1990) 10553.
%%%%%%%%%%%%%%%%%%%%%%%%%%%%%%%%%%%%%%%%%%%%%%%%%%%%%%%%%%%%%
\bibitem{kawakami}
N. Kawakami and S.-K. Yang: 
Phys. Lett. A {\bf 148} (1990) 359;  
N. Kawakami and S.-K. Yang: Phys. Rev. Lett. {\bf 65} (1990) 2309.
%%J. Phys.: Condens. Matter {\bf 3} (1991) 5983. 
%%%%%%%%%%%%%%%%%%%%%%%%%%%%%%%%%%%%%%%%%%%%%%%%%%%%%%%%%%%%%
\bibitem{kikoin}
K. Kikoin and Y. Avishi, 
cond-mat/0002444
%%%%%%%%%%%%%%%%%%%%%%%%%%%%%%%%%%%%%%%%%%%%%%%%%%%%%%%%%%%%
\bibitem{fujii2}
T. Fujii, A. Furusaki, N. Kawakami and M. Sigrist, 
in preparation
%%%%%%%%%%%%%%%%%%%%%%%%%%%%%%%%%%%%%%%%%%%%%%%%%%%%%%%%%%%%%
\end{thebibliography}
\end{document}